\documentclass[twocolumn]{aa}
\usepackage[dvips]{graphicx}
\usepackage{txfonts}
\begin{document}
\title{The mass-period distribution of close-in exoplanets}

\author{P. Ben\'itez-Llambay$^1$ \and F. Masset$^2$ \and C. Beaug\'e$^1$}

\institute{Observatorio Astron\'omico, Universidad Nacional de C\'ordoba, Argentina \and Instituto de Ciencias F\'\i sicas, Universidad Nacional Aut\'onoma de M\'exico, Apdo Postal 48-3, 62251 Cuernavaca, Mor., M\'exico\footnote{On leave from Service d'Astrophysique, CEA/Saclay, 91191 Gif-sur-Yvette, France }}

\abstract{The lower limit to the distribution of orbital periods $P$ for the current population of close-in exoplanets shows a distinctive discontinuity located at approximately one Jovian mass. Most smaller planets have orbital periods longer than $P \sim 2.5$ days, while higher masses are found down to $P \sim 1$ day.}  
{We analyze whether this observed mass-period distribution could be explained in terms of the combined effects of stellar tides and the interactions of planets with an inner cavity in the gaseous disk.}  {We performed a series of hydrodynamical simulations of the evolution of single-planet systems in a gaseous disk with an inner cavity mimicking the inner boundary of the disk. The subsequent tidal evolution is analyzed assuming that orbital eccentricities are small and stellar tides are dominant.}  
{We find that most of the close-in exoplanet population is consistent with an inner edge of the protoplanetary disk being located at approximately $P \gtrsim 2$ days for solar-type stars, in addition to orbital decay having been caused by stellar tides with a specific tidal parameter on the order of $Q'_* \simeq 10^7$. The data is broadly consistent with planets more massive than one Jupiter mass undergoing type II migration, crossing the gap, and finally halting at the interior 2/1 mean-motion resonance with the disk edge. Smaller planets do not open a gap in the disk and remain trapped in the cavity edge. CoRoT-7b appears detached from the remaining exoplanet population, apparently requiring additional evolutionary effects to explain its current mass and semimajor axis.}{}


\maketitle

\date

\section{Introduction}

Close-in planets (semimajor axis $a < 0.1$ AU) constitute a special subset of the exoplanetary population. Since it is unclear whether in-situ formation occurs, the current orbital and physical characteristics of these planets provide important constraints on their past evolution and formation process. Several mechanisms have been proposed to explain the pile-up of hot planets with a three~day orbital period, including a truncation of the gaseous disk by the star (Lin et al. 1996, Kuchner and Lecar 2002), planetary scattering combined with Kozai resonance and tidal circularization (Nagasawa et al. 2008), planetary evaporation (Davis and Wheatly 2009), and tidal interactions with the parent star (Jackson et al. 2009).
 
In particular, Kuchner and Lecar (2002) suggested that a giant planet in circular orbit could halt its migration when its orbital period was half that of the inner edge of the disk. In this configuration, all the planet's circular Lindblad resonances would lie in the inner cavity (IC) and no further interchange of angular momentum would take place. Masset et al. (2006) performed a series of hydrodynamical simulations to follow the evolution of low-mass planets in disks including an IC. They found that all bodies migrated until reaching a point slightly exterior to the cavity edge, where they were effectively trapped in a stable configuration in almost circular orbits. Although this result appears different from that predicted by Kuchner and Lecar (2002), each is valid, as we shall see, for a different range of planetary masses. 

The first reference to a possible correlation between mass and orbital period for close-in planets was proposed by Mazeh et al. (2005) for only six transiting bodies. They found that both parameters seemed to follow a linear law, with more massive bodies being located at smaller semimajor axes. Southworth et al. (2007) and Davis and Wheatley (2009) extended the analysis to a larger transiting population, finding a similar result although with a much broader dispersion. They proposed that smaller planets closer to the star might have been lost because of evaporation, similar to that currently ongoing at least in HD209458b (Vidal-Madjar et al. 2003) and WASP-17 (Anderson et al. 2010).

\begin{figure}
\centerline{\includegraphics*[width=20pc]{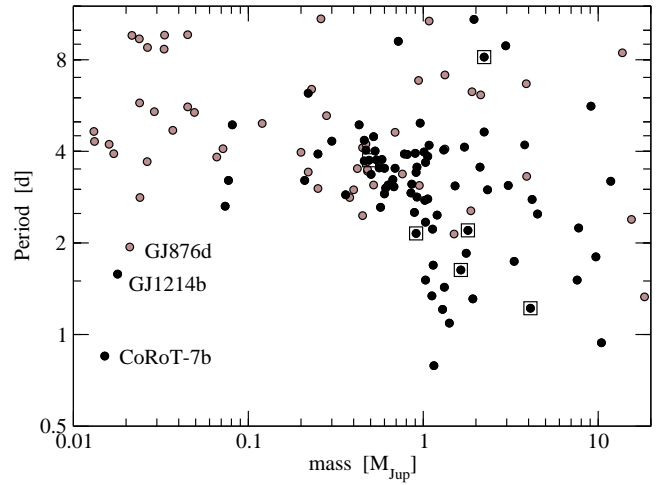}}
\caption{Distribution of orbital periods and planetary masses for close-in exoplanets with orbital period $P < 12$ days. Data from http://exoplanets.eu. Black circles denote planets with both Doppler and transit data, while gray circles mark bodies without detected transits. Empty squares correspond to planets with retrograde orbits with respect to the stellar spin.}
\label{fig1}
\end{figure}

Jackson et al. (2009) also analyzed the distribution of close-in planets, this time focusing on the correlation between the semimajor axis of the planet and the age of its star. They found that the lower limit of the semimajor axis was lower for younger stars, which implies that tidal effects could be responsible. Exoplanets with very short orbital periods in older stars would have had enough time to be tidally disrupted, thus they would only be presently observable in relatively young systems. 

In this paper, we revisit the mass-period distribution, taking advantage of the recent increase in the exoplanet population. Figure \ref{fig1} shows the orbital periods~$P$, as a function of the mass~$m$, for the known population of exoplanets with $P \le 12$~days (137 planets). Black circles correspond to cases for which both transits and Doppler data are available; bodies without detected transits are shown in gray. Exoplanets in apparent retrograde motion with respect to stellar rotation are identified by an empty square. These are WASP-8b (Queloz et al. 2010), WASP-17b (Anderson et al. 2010), WASP-33b (Collier Cameron 2010), Hat-P-7b (Winn et al. 2009), WASP-2b (Triaud et al. 2010), and WASP-15b and WASP-17b (Triaud et al. 2010). Although it may be argued that these bodies are not consistent with planetary migration (Triaud et al. 2010), they may also point towards primordial spin-orbit misalignment and not be related to subsequent orbital evolution of the planets (Lai et al. 2010).

The distribution exhibits a noticeable ``step'', exoplanets larger than one Jupiter mass ($M_{\rm Jup}$) appear to have a lower inner boundary (down to $\sim 1$ day) while for $m < M_{\rm Jup}$ the distribution seems restricted to larger values of $P$. The only exceptions are three bodies in the Super-Earth range, CoRoT-7b, GJ1214b, and GJ876d, which are all  marked in the plot. Of these, the latter two planets belong to low-mass stars ($M_*=0.17$ and $M_*=0.32$ solar masses, respectively), thus constitute special cases. CoRoT-7b, however, belongs to a solar-type star (Rouan et al. 2009). This planet has a very short orbital period ($\sim 0.85$ days) but also a very low mass ($m \sim 0.015 M_{\rm Jup}$), and does not seem to comply with the rest of the exoplanet distribution. In particular, Jackson et al. (2009) pointed out that CoRoT-7b could reach the Roche radius on timescales of $10^7 - 10^9$ years, depending on the value of the specific tidal parameter $Q'_*$. 

Regardless of these isolated cases, there seems to be a very clear discontinuity (or bump) in the mass-period distribution, located at approximately $m=M_{\rm Jup}$. Moreover, for high masses the lower limit in orbital periods appears very close to a $2/1$ mean-motion resonance with the disk edge for small planetary bodies. This appears consistent with a scenario in which the planetary traps proposed by Masset et al. (2006) would dominate the low-mass region, while the mechanism of Kuchner and Lecar (2002) would be mainly responsible for the upper end of the mass spectrum.

The main objective of this study is to test whether the combined action of planetary traps in the gaseous disk plus subsequent tidal interactions with the parent star could explain the observed distribution of close-in exoplanets. Since exoplanets in retrograde orbits should have exotic disk-planet and tidal evolutions, the study of these exosystems is beyond the scope of the present work, and we focus mainly on bodies believed to have orbital motion in the same direction as the stellar spin. Even in this case, we assume a zero inclination with respect to the stellar equator. 

In Section 2, we present a series of hydrodynamical simulations adopting different planetary masses and analyzing the relative halting distance from the central star. Not only are we interested in seeing whether such a hybrid and mass-selective process is possible, but also whether the boundary between both mechanisms is consistent with the observed distribution of close-in planets. Section 3 is devoted to the subsequent evolution of exoplanets under the stellar tide and their effects on any initial disk-driven distribution in the mass-period diagram. In Section 4, we analyze the case of the CoRoT-7 planetary system and present possible explanations of the present location of CoRoT-7b. Finally, conclusions close the paper in Section 5.

\begin{figure}
\centerline{\includegraphics*[width=20pc]{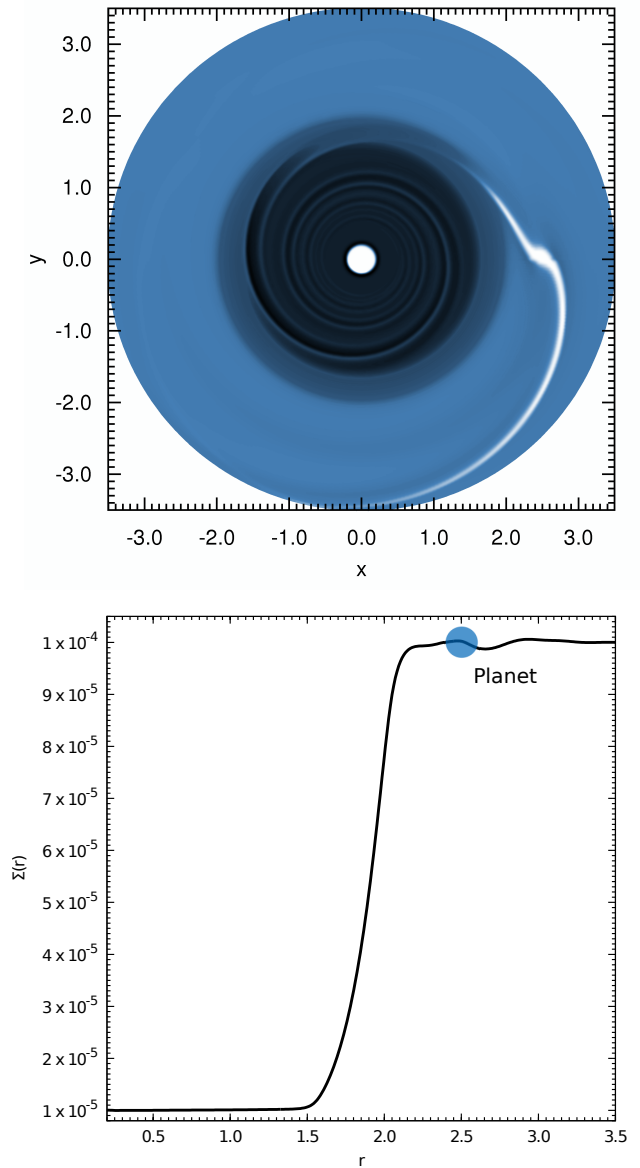}}
\caption{Snapshot of the surface density profile of one of our simulations. The inner cavity is centered around $r=1.8$. In the top frame, the low-density regions are shown in black, while the high-density regions are shown in white. The planet ($m = 0.1 M_{\rm Jup}$) is located on the $x$-axis.} 
\label{fig2}
\end{figure}

\section{Hydrodynamical simulations}

\subsection{Initial conditions}

Our simulations were carried out using the FARGO\footnote{See \tt http://fargo.in2p3.fr} hydro-code (Masset 2000a,b) with a central star of one solar mass. We consider two-dimensional, locally isothermal disks. The unit of length, $r_0$, is arbitrary (our simulations are scale free and can be rescaled to disks of arbitrary sizes), but it can be considered, for our purpose, that $r_0 \sim 0.01$~AU. Our disks, prior to a 1D relaxation dedicated to the creation of the IC, have a uniform surface density $\Sigma=10^{-4}\;M_\odot.r_0^{-2}$, which translates, in our case, to $\Sigma\sim900$~g.cm$^{-2}$. The asymptotic radius of our planets is virtually independent of this value, which essentially prescribes how long it takes for them to reach it. In a similar manner, the kinematic viscosity of our disks is uniform outside the cavity, and is set to $\nu=10^{-5}r_0^2\Omega_0^{-1}$, where $\Omega_0$ is the orbital frequency at radius $r_0$. Some initial test runs were performed with other values of $\Sigma$ and $\nu$ with no significant change in the results, except for the migration timescale. The disk aspect ratio is set uniformly to $H/r=0.05$, hence the value of the $\alpha$ parameter that characterizes turbulence (Shakura \& Sunyaev, 1973) just outside the IC (see below) is $\alpha \sim 3\cdot 10^{-3}$. 

\begin{figure}
\centerline{\includegraphics*[width=20pc]{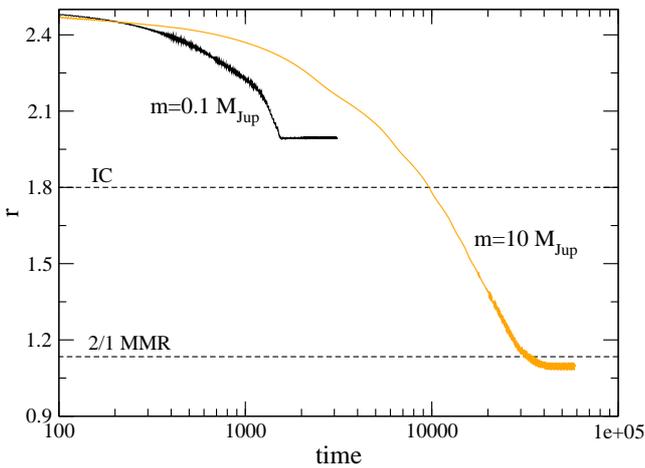}}
\caption{Evolution of semimajor axis of two one-planet systems migrating and halting near the disk inner cavity edge (IC). Black curve corresponds to $m=0.1 M_{\rm Jup}$ and gray to $m=10M_{\rm Jup}$. The IC is centered at $r_{IC}=1.8$, while its interior $2/1$ mean-motion resonance (MMR) is located at $r \simeq 1.13$. Time is in orbital periods.}
\label{fig3}
\end{figure}

Our disks are described on a polar grid with $384$~zones in azimuth and $306$~zones in radius. The inner radius of the mesh is at $0.2r_0$, while the outer radius is at $3.5r_0$. The IC was generated using an {\em ad hoc} step in kinematic viscosity around $r=1.8r_0$, using the same recipe as described in Masset et al. (2006), adopting a value of $F = \Sigma_o/\Sigma_i = 10$ for the ratio of the surface densities outside and inside the cavity. We obtain a rather sharp IC of width $\Delta r \simeq 0.5r_0$. Boundary conditions were chosen to be non-reflecting for the inner edge and such that there was a continuous outer source mass to maintain the surface density in the external regions of the disk (Masset et al. 2006 used non-reflecting boundary conditions for both edges). No planet was considered at this point. 

An initial one-dimensional run was performed to allow the cavity to form and reach a steady state configuration. The resulting density profile and total torque were analogous to those shown in Masset et al. (2006), showing the existence of a stable fixed point in the torque at a distance $r^*$ slightly larger than the nominal edge of the IC (Figure \ref{fig2}). A planet located outside $r^*$ would feel a negative torque and suffer a negative orbital decay towards the star. Conversely, a body placed at a radial distance slightly smaller than $r^*$ would feel a positive torque and an outward radial migration. Thus, the point $r=r^*$ is a stable nesting place for the planet, as long as its mass is sufficiently low to avoid disruption of the density (and torque) profile. 

\subsection{Test simulations of planet trapping}

After the disk density profile had stabilized, we included a planet of mass $m$ in an initially circular orbit at $r=2.5r_0$. The system was allowed to evolve until the planet reached a stationary solution and the orbital decay effectively stopped. Figure \ref{fig3} shows the results of two runs, the first with $m = 0.1 M_{\rm Jup}$ and the second with $m = 10 M_{\rm Jup}$. The location of the center of the IC is marked by the top horizontal dashed line, while the lower corresponds to the interior $2/1$ mean-motion resonance (MMR) with the center of the IC. 

The smaller planet suffers an almost constant orbital decay until it is halted at a location slightly outside the cavity, in accordance with the findings of Masset et al (2006). The more massive planet, however, suffers a Type II migration that opens a gap in its co-orbital region and completely disrupts the density profile of the disk in its vicinity. The cavity edge is therefore incapable of generating a strong corotation torque and fails to trap the planet, which continues its orbital decay inside the cavity. The migration is finally stopped very close to the location of the $2/1$ MMR with the IC where the differential Lindblad torque reduces to zero, as predicted by Kuchner and Lecar (2002). 

In all cases, the eccentricities were damped, and the final orbits were observed to be quasi-circular with $e \sim 10^{-3}$. This final eccentricity is different from the results presented by Rice et al. (2008), where a significant eccentricity excitation is observed for Type II migration inside the inner disk edge. From what we have been able to see, the discrepancy arises from the different numerical setup. Rice et al (2008) truncate the disk at the mesh's inner edge, and the planets that they consider essentially orbit in a vacuum. This is in contrast to our situation in which some residual material surrounds the orbit. This material, by the action of co-orbital Lindblad resonances, is a powerful source of eccentricity damping.

\subsection{Scaling and comparison with observations}

\begin{figure}
\centerline{\includegraphics*[width=20pc]{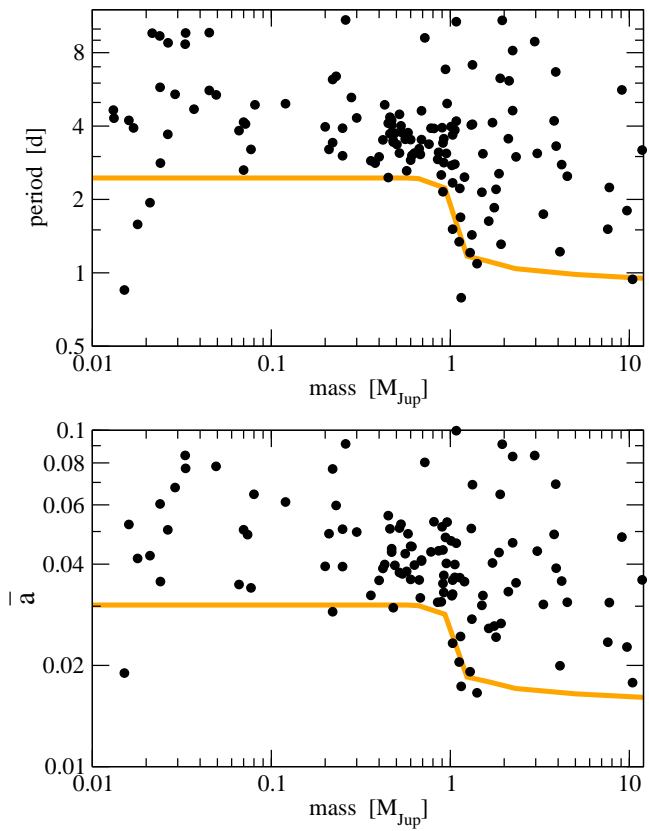}}
\caption{Gray circles reproduce the mass-period distribution of known close-in exoplanets. The broad orange band represents the final semimajor obtained from the runs, scaled along the $y$-axis to fit the observational data. In the lower frame, ${\bar a}$ denotes the normalized semimajor axis defined in Eq. (\ref{eq5}).}
\label{fig4}
\end{figure}

We repeated the simulations for a total of $18$~runs, with planetary masses ranging from $0.01$ to $20$ Jupiter masses, all starting from initial circular orbits at $r=2.5r_0$. The orange curve in the top plot of Figure~\ref{fig4} presents the final orbital period as function of the planetary mass. For comparison, the observational data are again drawn in circles. 

Since the location of our IC was in arbitrary units, we have a degree of freedom in the sense that we may shift the orange curve arbitrarily in the $y$-axis to fit the observed exoplanets. We find that the best fit gives an inner cavity edge located at approximately $a \sim 0.03$ AU corresponding to an orbital period of $\sim 2$ days. This value is comparable with the inferred inner gas radii of T Tauri star disks from CO spectroscopy (Najita et al. 2007, Carr 2007) which seems to be located around $0.03-0.04$~AU. Although our best-fit distance is slightly lower, any difference could be due to stellar parameters and the maximum CO velocity (Carr 2007). 

Exoplanets with masses below one Jovian mass seem to have a lower-limit orbital period of about $2-3$~days, consistent with the position of the equilibrium point being slightly beyond the IC (Masset et al. 2006). On the other hand, the orbital periods of more massive planets have a lower limit, this time in accordance with the location of the $2/1$ mean-motion resonance with the IC, as predicted by Kuchner \& Lecar (2002). 

\section{Tidal evolution of close-in exoplanets}
\label{sec:tidal-evol-close}
Although the results presented in the top graph of Figure \ref{fig4} appear to be encouraging, there are two important approximations that must be examined. First, the exoplanets orbit stars with different radii and masses, and we have assumed that the scaling in the $y$-axis is the same. Although we have not assumed any origin for the inner cavity (MRI, stellar wind, etc.), it is almost certain that the location of the IC should be a function of the stellar mass. If we assume a very simple model in which $r_{IC}$ scales with $R_*$, then the position of both GJ1214b and GJ876d would be displaced above the broad orange curve, thus eliminating their incongruity with respect to the simulations. However, CoRoT-7 has a stellar mass of $0.93 M_\odot$ and its location in the diagram would still be conflicting.

A second approximation is that we have neglected the later evolution of the planetary periods caused by stellar tides. Assuming almost circular planar orbits for the close-in planets, we can neglect the planetary tides and approximate the differential equation for orbital decay as
\begin{equation}
\label{eq1}
\frac{da}{dt} = -\biggl[ \frac{9}{2} \left( \frac{G}{M_*} \right)^{1/2} \frac{R_*^5 M_p}{Q'_*} \biggr]
                \; a^{-11/2}
\end{equation}
(see Ferraz-Mello et al. 2008, Jackson et al. 2009). This expression can be easily integrated to yield
\begin{equation}
\label{eq2}
a(t)^{13/2} = \biggl[ \frac{117}{4} \left( \frac{G}{M_*} \right)^{1/2} \frac{R_*^5 M_p}{Q'_*} \biggr] t + a_0^{13/2} ,
\end{equation}
where $a_0=a(t=0)$ is the initial value of the semimajor axis. After some simple algebraic manipulations, we can express this solution as
\begin{equation}
\label{eq3}
{\bar a}(t)^{13/2} = {\bar a}_0^{13/2} - \beta t ,
\end{equation}
where
\begin{equation}
\label{eq4}
\beta = \frac{117}{4} \biggl(\frac{G}{M_\odot}\biggr)^{1/2} {R_\odot}^5 \frac{M_p}{Q'_*} 
\end{equation}
and 
\begin{equation}
\label{eq5}
{\bar a} = a \; \biggl(\frac{M_*}{M_\odot}\biggr)^{1/13} \biggl(\frac{R_\odot}{R_*}\biggr)^{10/13} .
\end{equation}
Equation (\ref{eq3}) is then independent of the stellar parameters, whose values are incorporated into the ``normalized'' semimajor axis ${\bar a}$. We note that the rate of orbital decay is given by $\beta$, which is linearly proportional to the planetary mass and inversely proportional to the stellar dissipation parameter $Q'_*$. Is it of course probable that the tidal parameter itself depends on the stellar parameters (see Barker and Ogilvie 2009), but there has so far been no clear indication of how it may vary nor by what magnitude.

The lower plot in Figure \ref{fig4} shows the distribution of ${\bar a}$ as a function of $m$ for the close-in exoplanet population. The orange curve marks the position of the simulated lower limit to the semimajor axis. As expected, both GJ876d and GJ1214b are now above the lower limit, although CoRoT-7b still remains detached. We note that the bump in the distribution is still clearly visible.

We can now use equation (\ref{eq3}) to check whether the present-day distribution could originate solely from tidal evolution. To test this idea, we generated a synthetic population of $10^4$ fictitious exoplanets distributed randomly in planetary mass across the interval $m \in [0.01,12] M_{\rm Jup}$ and with random stellar ages $T$ between one and eight Gyr. The normalized semimajor axis ${\bar a}$ of each planet was then evolved tidally for the age of the system using a stellar tidal parameter $Q'_*=10^6$. 

\begin{figure}
\centerline{\includegraphics*[width=20pc]{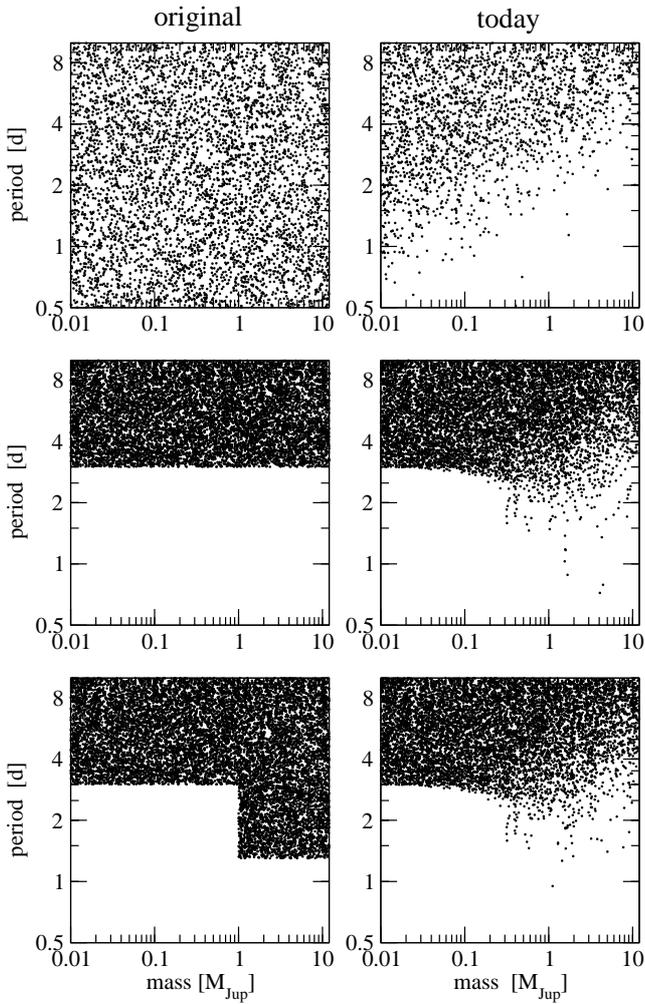}}
\caption{Tidal evolution of fictitious planets (stellar tides alone) with $Q'_*=10^6$. Initial distributions are shown on the left, while the evolved distributions are shown on the right. Stellar ages where chosen uniformly between one and eight Gyr.}
\label{fig5}
\end{figure}

\begin{figure}
\centerline{\includegraphics*[width=20pc]{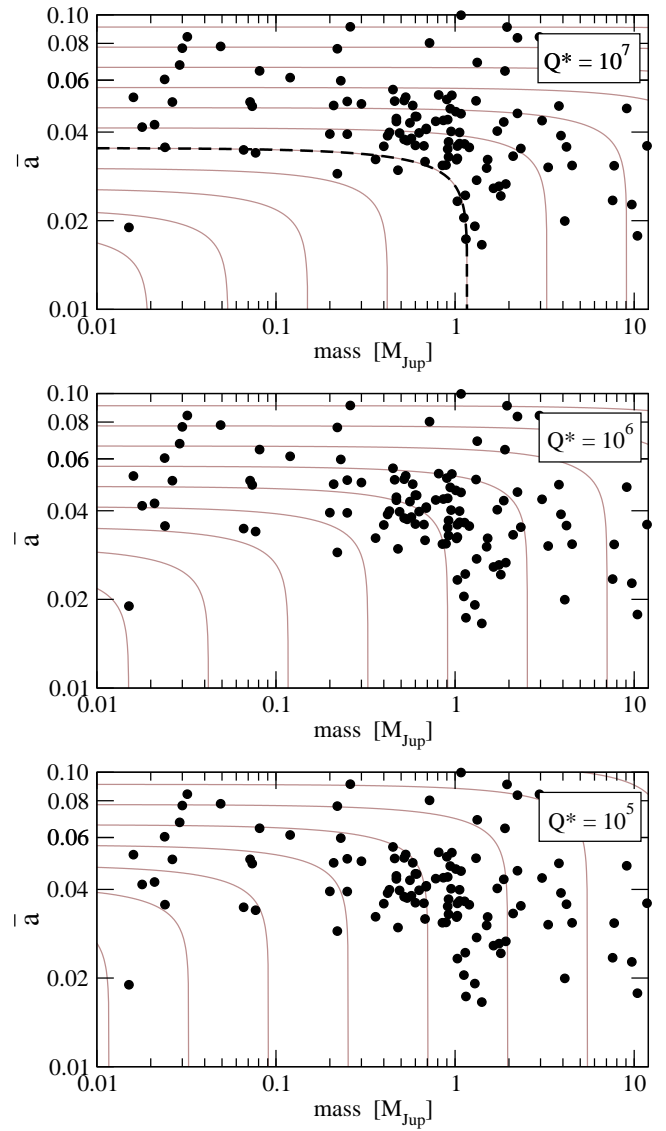}}
\caption{Distribution of close-in exoplanets in the ${\bar a}$-$m$ diagram. Gray curves show the end evolution due to stellar tides (at $T=8$ Gyr) of initial constant values of ${\bar a}$. Each curve corresponds to a different ${\bar a}_0$. Stellar tidal parameters are given in the top right-hand side of each graph. The broad dashed curve in the top panel, corresponding to ${\bar a}_0=0.035$, gives the best fit to the lower limit of the current exoplanet population.} 
\label{fig6}
\end{figure}

Figure \ref{fig5} shows results considering three different initial distributions, chosen to be uniform in orbital period $P$ and $\log{(m)}$. In all cases, the lowest mass was equal to $m = 0.01 M_{\rm Jup}$, and the largest to ten Jovian masses. Since this preliminary analysis is only intended to be illustrative, we assumed solar-type stars for all bodies; consequently, the normalized semimajor axis ${\bar a}$ is equal to the nominal semimajor axis $a$. 

For the top plots, we assumed fictitious planets with orbital periods that have no lower limit. The initial data set is shown in the left plot, while the right-hand plot presents the final distribution after evolution through stellar tides for $T=8$ Gyr. Since the decay rate $\beta$ is proportional to the planetary mass, more massive planets fall more rapidly towards the star, leading to a final population that has a lack of large bodies with small orbital periods. This distribution shows little relation to the real distribution of planets. 

For the middle plots, we considered a lower limit to the orbital periods equal to $3$ days. This value is sufficiently large for terrestrial-type bodies to be virtually unaffected, although giant planets still suffer significant orbital decay. Depending on their initial semimajor axis and stellar age, many massive planets reach the Roche radius and are engulfed by the star, but a portion of the population remains. We note the final step-like distribution, which is reminiscent of the real population. However, the bump is now located at $\sim 0.3 M_{\rm Jup}$, thus at a lower mass than for the real exoplanets.

Finally, in the lower plots of Figure \ref{fig5} we considered an initial population with a step in the orbital period, similar to that resulting from the hydrodynamical simulations. Even though the initial conditions are different from those depicted in the middle plots, there is no significant difference in the final distribution. This seems to indicate that any evidence of an initial disk-related structure would be smeared by the later tidal evolution. Thus, a bump in the present real population is not necessarily indicative of a dynamical structure prior to the dissipation of the gaseous disk.

These results indicate the clear possibility that the observed bump in the exoplanet ${\bar a}$-$m$ distribution could be mainly due to stellar tidal effects. To test this proposal in more detail, we calculated how a lower bound in the values of ${\bar a}$ would be modified, for different values of $Q'_*$, after a timescale of $T=8$ Gyr. Results are shown in Figure \ref{fig6} for three values of the stellar tidal parameter. Each gray curve shows the function
\begin{equation}
\label{eq6}
{\bar a} = \biggl( {\bar a}_0^{13/2} - \beta T \biggr)^{2/13} \hspace*{1cm} {\rm with \;\;\;} T=8 {\rm Gyr}
\end{equation}
for different values of ${\bar a}_0$. For comparison, each plot also reproduces the present distribution of close-in exoplanets. 

The present-day lower limit to the normalized semimajor axis shows a very good agreement assuming ${\bar a}_0 \simeq 0.035$ and a tidal evolution with $Q'_*=10^7$, represented by a bold dashed curve in the top graph. This value of the normalized semimajor axis corresponds to original orbital periods of $P \sim 2.5$ days for solar-type stars. We note that the other plots, corresponding to smaller values of $Q'_*$ do not show good correspondence for any adopted value of ${\bar a}_0$. This result appears to indicate that smaller values of $Q'_*$, and consequently faster orbital decays due to stellar tides, are inconsistent with the current population of close-in exoplanets.

\begin{figure}
\centerline{\includegraphics*[width=20pc]{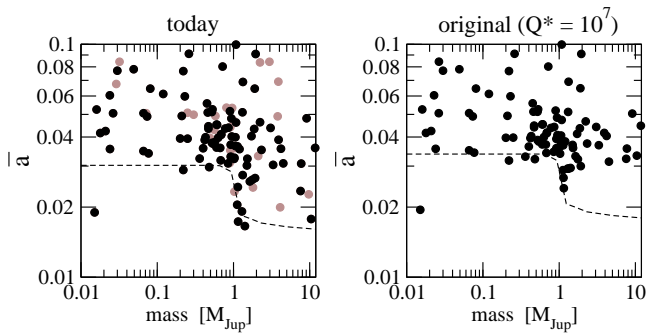}}
\caption{{\bf Left:} Current-day distribution of exoplanets in the ${\bar a}$-$m$ diagram. Black circles show planets for which values of $M_*$, $R_*$ and stellar ages $T_*$ are available. Gray circles show those bodies for which stellar
data is missing. {\bf Right:} Reconstructed original distribution assuming backward orbital evolution from stellar tides for the age of each parent star. In both plots, the dashed curves show the lower limit to normalized semimajor axis, as a function of the planetary mass, obtained from the hydrodynamical simulations and scaled along the $y$-axis to fit the exoplanets.}
\label{fig7}
\end{figure}

\begin{figure}
\centerline{\includegraphics*[width=20pc]{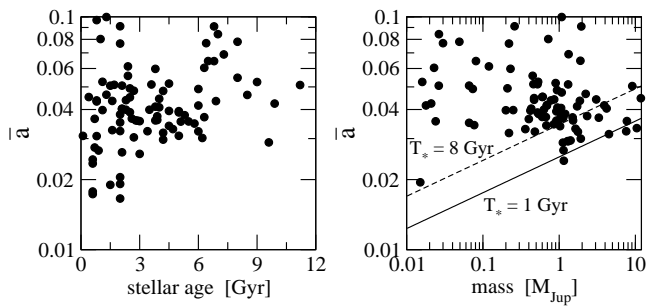}}
\caption{{\bf Left:} Current values of the normalized semimajor axis ${\bar a}$ for close-in exoplanets as a function of the stellar age $T_*$. Notice that bodies with ${\bar a} < 0.03$ are limited to young systems ($T_* < 3 Gyr$), while larger semimajor axes are found for all stellar ages. {\bf Right:} Original distribution 
in the $m$-${\bar a}$ diagram. Diagonal curves give limits to the orbital decay leading to disruption for two stellar ages: the continuous line show results for
$T_*=1$ Gyr, while the dashed line corresponds to $T_*=8$ Gyr. }
\label{fig8}
\end{figure}

Although the timescale ($T=8 {\rm Gyr}$) chosen for Figure \ref{fig6} may seem arbitrary, equations (\ref{eq4}) and (\ref{eq5}) show that the most relevant parameter for the orbital evolution is actually the ratio $T/Q'_*$. Since the uncertainty in the tidal parameter is much larger than in the age of the system, it seems justified to use a single fixed value of $T$ and to assume that no qualitative differences would be observed for other stellar ages. 

As a final test, we can attempt to reconstruct the original mass/semimajor-axis distribution of the exoplanet population, integrating each current value of ${\bar a}$ backwards in time using equation (\ref{eq6}), adopting in this case the value of $T$ equal to the age of each parent star $T_*$. To perform this calculation, we require in addition to the stellar mass $M_*$ and radius $R_*$, estimates of each stellar age $T_*$. However, this information is not available for all planetary systems. Out of the 133 original planets from our data set in direct orbits (i.e. eliminating the bodies believed to be in retrograde motion), we were only able to obtain a full set of stellar parameters for 94 exoplanets. Stellar data were obtained from the Simbad database for stellar properties and from Jackson et al. (2009). 

The present-day ${\bar a}$-$m$ distribution for this reduced population is shown as black circles in the left-hand plot of Figure \ref{fig7}. For comparison, gray circles show those planets for which complete stellar data is currently unavailable. The location of the lower bound to the semimajor axis deduced from the hydrosimulations is shown as a bold dashed curve. Unfortunately, many systems with giant planets with ${\bar a} < 0.025$ do not have complete stellar properties, and the bump in the distribution is not clearly visible for the smaller population. 

The right-hand side plot of Figure \ref{fig7} shows the ``original'' location of the smaller population after the backwards integration for each stellar age. For comparison, the inner-cavity-induced lower limit is again shown as a bold dashed 
curve, although the scaling in the $y$-axis has been modified to fit the new values of semimajor axis. Although the distribution still shows a good agreement for planetary masses up to $m \sim 1 M_{\rm Jup}$, most of the higher masses have increased their semimajor axis beyond ${\bar a} \sim 0.03$, and little evidence remains of the bump. 

A possible explanation could be that most giant planets with small values of ${\bar a}$ were subsequently lost due to tidal disruption. To check this hypothesis, the left plot of Figure \ref{fig8} shows the relation between the current values of the normalized semimajor axis as a function of the stellar age. Although planets with relatively large semimajor axes exist for all values of $T_*$, it seems very clear that giant planets with ${\bar a} < 0.025$, which are primarily responsible for the bump in the distribution, belong to young systems with a maximum stellar age of $T_* \simeq 3$ Gyr. 

From Eq. (\ref{eq3}), it is possible to estimate, for each planetary mass, the critical value of the normalized semimajor axis (i.e. ${\bar a}_0^*$) that falls towards the star of a given stellar age $T_*$. This is given approximately by
\begin{equation}
\label{eq7}
{\bar a}_0^* = (\beta T_*)^{2/13} , 
\end{equation}
which is simply obtained by setting the final semimajor axis to be equal to zero. Although the planet is believed to disrupt upon reaching the Roche radius and, thus, before impacting the star itself, as shown by Jackson et al. (2009) the orbital decay at such small semimajor axes is so swift that the timescales for both scenarios are practically equal. 

The right-hand plot of Figure \ref{fig8} once again reproduces the ``original'' distribution of exoplanets in the $m$-${\bar a}$ diagram shown before in Figure \ref{fig7}. The diagonal lines show the values of ${\bar a}_0^*$, as a function of the planetary mass, for two stellar ages: $T_*=1$ Gyr (continuous line) and $T_*=8$ Gyr (dashed line). It is clear that even if giant planets were deposited by a hypothetical disk inner cavity at small values of the semimajor axis (${\bar a}<0.025$), they would be rapidly absorbed by the parent star because of tidal effects, even for planetary systems as young as $1$ Gyr. Thus, it is not unexpected that if any original bump in the distribution were created by disk-planet interactions, subsequent tidal effects would have eliminated most traces.

\section{CoRoT-7b}

CoRoT-7 appears to be a special case. As seen from the lower plot of Figure \ref{fig4}, its low mass and short orbital period mean that it is well separated from the $m$-${\bar a}$ distribution observed for other close-in exoplanets. We recall that this planet orbits a solar-type star of mass $M_* \simeq 0.93 M_\odot$.  

Jackson et al. (2009) proposed that these planetary bodies should be undergoing significant tidal evolution and orbital decay. Most bodies in this mass range would then be absorbed by the star on timescales shorter than the age of the star. According to this idea, CoRoT-7b owes its present existence solely to its star being relatively young ($\sim 1.5$ Gyr). 

However, numerical simulations of the tidal evolution of the CoRoT-7 planetary system (Ferraz-Mello et al. 2010) indicate that the current eccentricities should be extremely low ($e \ll 10^{-3}$), and that any primordial departure from circular motion would have been rapidly damped before any significant orbital decay occurred. Thus, tidal evolution would have been given primarily by stellar tides alone. In this case, the original location of CoRoT-7b in the primordial $m$-${\bar a}$ distribution should be given by the right-hand plots of Figures \ref{fig7} and \ref{fig8}; once again this planet appears to be detached from the remaining close-in planetary population. 

Since CoRoT7 harbors at least one additional planet, it is possible that mutual dynamical interactions might also explain this planet's proximity to the star. In multiple-planet systems, scattering is believed to have played an important role in sculpting the general exoplanet distribution, and the same phenomena may have occurred in this system. According to this idea, CoRoT7-b could have had a close encounter with CoRoT7-c (or with an additional ejected planet) and suffered a significant reduction in its semimajor axis. Subsequent tidal interactions would have circularized its orbit to its present state. However, CoRoT7 is not the only multiple-planet system in this region. Both the GI581 and HD40307 planetary systems have two known super-Earths with short orbital periods that nevertheless lie above the expected lower limit. Hence once again CoRoT7 seems to be different. 

A possible explanation may lie elsewhere. Valencia et al. (2010) and Jackson et al. (2010) proposed that CoRoT-7b could be the solid core of a primordial giant planet whose gaseous envelope was lost due to evaporation. A recent re-evaluation of the radial velocity data by several authors indicate that the mass of CoRoT-7b could be as high as $9 M_\oplus$ (Ferraz-Mello et al. 2010) or as low as $2 M_\oplus$ (Pont et al. 2010). Although a high current mass appears indicative of a rocky/iron composition and that the original mass was at most twice the present value (Valencia et al. 2010), a lower mass is consistent with a lighter composition and a envelope-depleted gas giant.

If CoRoT-7b were indeed the solid core of a primordial gas giant, then its location in the $m$-${\bar a}$ diagram (right-hand plot of Figure \ref{fig7}) would be located close to the dashed curve and, thus, consistent with the rest of the planetary population. We still need of course to explain how this planet suffered an evaporation of its gaseous envelope on timescales shorter than one Gyr, and why it appears to be the only example of this effect.

\section{Conclusions}

We have attempted to understand the dynamical origin and evolution of the mass-period distribution of close-in exoplanets. The present-day population shows a distinctive discontinuity located at approximately one Jovian mass. Smaller planets have orbital periods longer than $P \sim 2.5$ days, while higher masses are found to have periods as short as $P \sim 1$ day. 

We have found that the combined effects of tidal evolution and disk-planet interactions with an inner cavity (IC) in the gaseous disk can explain most of the observed characteristics. The current distribution appears to be compatible with an inner disk edge located approximately at distances of ${\bar a} \simeq 0.035$, which for solar-type stars corresponds roughly to orbital periods of $P \simeq 2.5$ days. This value is consistent with the inner gas radii for T Tauri stars as estimated from CO spectroscopy (Najita et al. 2007, Carr 2007). 

Planets below a certain critical mass $m_c \sim M_{\rm Jup}$ are trapped just outside the IC as found by Masset et al. (2006). The location of the stationary solution with respect to the IC is practically mass-independent. In contrast, bodies with $m > m_c$ enter a regime characterized by a Type II migration that causes significant perturbations to the density profile of the disk; consequently  the IC edge cannot generate a significant positive corotational torque and does not stop the orbital decay. As predicted by Kuchner and Lecar (2002), migration only brakes inside the inner edge at a $2/1$ mean-motion resonance with the cavity edge. 

For reasonable values of the disk viscosity, we expect a gap opening to occur when the height of the disk is approximately equal to the Hill radius of the planet. For a scale height equal to $H/r \simeq 0.05$, this implies a minimum mass of $m \simeq 0.4 M_{\rm Jup}$, a value similar to our critical mass $m_c$. Adopting a value of one Jovian mass for the critical mass leads to a slightly larger value of $H/r \simeq 0.07$. However, given the uncertainties involved in both the gap opening criteria and its dependence on the viscosity, the values may be considered to be virtually equivalent. Thus, the location of the bump in the observed distribution of close-in planets appears to be consistent with a mass threshold for gap opening (Crida et al. 2006 and references therein). Although the present data are sparse and plagued by the additional effects described in this work, we may expect that the aforementioned results will be confirmed by further detections. In contrast, when the statistics become sufficiently robust, the location of this bump may be used to place constraints on the physical properties of the inner disk, in terms of both temperature and effective viscosity, as these quantities feature in the gap opening criterion.

Since the edge of the IC created in our simulations is placed at an arbitrary distance from the star, we have a certain degree of freedom when fitting the numerical mass-period distribution to the real planets (as seen in Figure \ref{fig4}). We have adopted a vertical displacement that minimizes the number of exoplanet inside the cavity edge, but this is not the only option. It may be argued that it would be better to fit the synthetic curve with the location of sub-giants (i.e. $m \sim 0.5 M_{\rm Jup}$) in the $m$-$P$ diagram. We note that these bodies exhibit a smaller dispersion in orbital period than observed for any other mass range, and can be seen as a compact group in both plots of Figure \ref{fig4}. However, this displacement would lead to our accepting a larger number of small planets inside the cavity edge, bodies whose subsequent tidal evolution should have been very small. 

Whatever the choice, the qualitative results are not affected. Moreover, the ratio of the stopping values of $P$ (for small and large bodies) is scale-independent, and found to be slightly larger than $2$. This is because although higher masses are stopped in a $2/1$ MMR with the IC, the small bodies are trapped outside the disk edge. Once again, the distribution of real planets seems to yield a similar ratio. 

The subsequent tidal evolution of the close-in planets is consistent with a stellar tidal parameter of $Q'_* = 10^7$, a value similar to that predicted by Schlaufman et al. (2010) from synthetic population models. Smaller parameters, leading to higher rates of orbital decay, do not lead to distributions similar to the observed population. This is also consistent with the analysis of Ogilvie \& Lin (2007). A consequence of the tidal evolution is the removal of most of the original gas giants with short orbital periods and their substitution by exoplanets that were originally farther away. Thus, we expect that many of the primordial planets with $P < 2$ days and $m > 1 M_{\rm Jup}$ might have been tidally disrupted and absorbed by their parent stars.

Although this scenario is consistent with the properties of most of the exoplanetary population, it appears difficult to explain the present-day mass and orbit of CoRoT-7b. A possible explanation is to assume that the planet started its life as a gas giant whose gas envelope was completely evaporated (Valencia et al. 2010, Jackson et al. 2010). 

Last of all, in the present scenario we have neglected the role of the orbital inclinations; however, the same results should be expected as long as the inclinations are not very large. Three-dimensional studies of disk-planet interactions in the linear approximation (e.g. Tanaka and Ward 2004), as well as numerical simulations (e.g. Cresswell et al. 2007), show little effect on a finite inclination on the migration timescale. Similar results are also found for orbital evolution due to tidal effects (e.g. Ferraz-Mello et al. 2008, Barker and Ogilvie 2009). For planets in retrograde orbits results are, however, is more difficult to evaluate. It is unclear whether gas disks in retrograde motion (with respect to the stellar spin) would have inner cavities or how such a structure would interact with planets in its vicinity. Similarly, the tidal evolution of bodies in retrograde motion is poorly understood, thus it	is impossible at present to ascertain how the results of this work could be extended to these systems.

\section*{Acknowledgments}
This work has been partially supported by the Argentinian Research Council -CONICET-. F.M. and C.B. would like to acknowledge the invitation to participate in the XII Brazilian Colloquium of Orbital Dynamics, where the original idea for this work was discussed. A substantial part of the work was developed during the program "Dynamics of Disks and Planets", held from August 15 to December 12, 2009 at the Newton's Institute of Mathematical Science at the University of Cambridge (UK). C.B. would like to thank the organizers of the program and to fruitful discussions with all the participating researchers. Most of the numerical simulations performed in this work have been run on a 140 core cluster funded by the program {\em Origine des Plan\`etes et de la Vie (OPV)} of the French {\em Institut National des Sciences de l'Univers (INSU)}.

\end{document}